\documentclass[preprint,showpacs,preprintnumbers,amsmath,amssymb]{revtex4}

\usepackage{graphicx}
\usepackage{dcolumn}
\usepackage{bm}

\begin{document}

\preprint{}

\title{Hot electron energy relaxation in lattice-matched InAlN/AlN/GaN heterostructures: the sum rules for electron-phonon interactions and hot-phonon effect}

\author{J.-Z. Zhang}
 \email{phyjzzhang@jlu.edu.cn}
 \affiliation{School of Physics, Jilin University, Changchun, 130012, China.
}
\author{A. Dyson}%
\affiliation{Department of Physics, University of Hull, Hull, HU6 7RX, UK.
}%

\author{B.~K. Ridley}
\affiliation{School of Computing Science and Electronic Engineering, University of Essex,
Colchester, CO4 3SQ,
UK.
}%

\date{\today}

\begin{abstract}
Using the dielectric continuum (DC) and three-dimensional phonon (3DP) models, energy relaxation of the hot electrons in the quasi-two-dimensional  channel of lattice-matched InAlN/AlN/GaN heterostructures is studied theoretically, taking into
account non-equilibrium polar optical phonons, electron degeneracy, and screening from the
mobile electrons.
The electron power dissipation and energy relaxation time due to both half-space and interface phonons are calculated as functions of the electron temperature $T_e$ using a variety of phonon lifetime values from experiment, and then compared with those evaluated by the 3DP model. Thereby particular attention is paid to examination of the 3DP model to use for the hot-electron relaxation study. The 3DP model yields very close results to the DC model: with no hot phonons or screening the power loss calculated from the 3DP model is 5\% smaller than the DC power dissipation, whereas slightly larger 3DP power loss (by less than 4\% with a phonon lifetime from 0.1 to 1 ps) is obtained throughout the electron temperature range from room temperature to 2500 K after including both the hot-phonon effect (HPE) and screening. Very close results are obtained also for energy relaxation time with the two phonon models (within a 5\% of deviation). However the 3DP model is found to underestimate the HPE by 9\%.
The Mori-Ando sum rule is restored by which it is proved that the power dissipation values obtained from the DC and 3DP models are in general different in the pure phonon emission process, except when scattering with interface phonons is sufficiently weak, or when the degenerate modes condition is imposed, which is also consistent with Register's scattering rate sum rule. The discrepancy between the DC and 3DP results is found to be caused by how much the high-energy interface phonons contribute to the energy relaxation: their contribution is enhanced  in the pure emission process but is dramatically reduced after including the HPE.
Our calculation with both phonon models has obtained a great fall in  energy relaxation time at low electron temperatures ($T_e<$ 750 K) and slow decrease at the high temperatures with the use of decreasing phonon lifetime with $T_e$. The calculated temperature dependence of the relaxation time and the high-temperature relaxation time $\sim$0.09 ps are in good agreement with experimental results. 
\end{abstract}

\maketitle

\section{Introduction}

Nitride compound semiconductors such as GaN and AlN which have a wide energy gap withstand high breakdown electric fields and support excellent thermal stability \cite{Morkoc}. 
The heterostructure field-effect transistor (HFET) using a GaN-based heterostructure is very promising for high-power radio frequency and high-mobility operations, owing to its high electron density, high electron drift velocity as well as its high drain breakdown voltage.
Particularly interesting are lattice-matched InAlN/AlN/GaN heterostructures \cite{Matulionis:2008,Matu09,Leach:2009,Leach:2010}, in which a two-dimensional electron gas forms in the undoped GaN layer near the interface of the GaN layer and the one nanometre thick AlN spacer,  due to internal {\it spontaneous polarization} alone, i.e., with no strain-induced polarization by the piezoelectric effect.  As such an electron gas arises in the absence of doping and lattice strain,  high electron density ( 2  $\times10^{13}$/cm$^2$) as well as high drift velocity (3 $\times 10^7$ cm/s) can occur in the lattice-matched heterostructures.
The electrons in the quasi-two-dimensional (quasi-2D) GaN channel are heated up due to high electric power,
with the electron temperature $T_e$ elevated above the lattice temperature $T_0$ (i.e. room temperature).
Electron temperatures up to 2500 K have been measured using a microwave noise technique for a lattice-matched InAlN/AlN/GaN heterostructure with an areal electron density of $1.2\times 10^{13}$/cm$^2$ in the GaN channel \cite{Matulionis:2008}.
Heat dissipation in the GaN channel is a complicated process which includes energy relaxation of the hot electrons mainly by emission of polar-optical phonons, decay of the polar-optical phonons into acoustic phonons via anharmonic interactions and diffusion of the excess acoustic phonons into the remote heat sink.
The optical phonon lifetime has been measured for GaN-based heterostructures, which in general falls in the range from 0.1 to 1.7  ps \cite{Matulionis:2008,Leach:2010,Liberis:2009} - except for the case of the electron density $8\times 10^{11}$/cm$^2$, and depends on both the electron density \cite{Liberis:2009} and electron temperature \cite{Matulionis:2008,Liberis:2009}.
As the optical phonon lifetime is much longer than the Fr{\"o}hlich scattering time ($\sim$10 fs for GaN \cite{Zhang:2014}), the emission of polar optical phonons is very fast compared to their decay into acoustic phonons, and a large population of nonequilibrium ("hot") phonons are accumulated leading to a slowdown in energy relaxation (termed hot-phonon effect). Hot phonons also result in an increase of electrical resistance and impose limitations on electron drift velocities \cite{Leach:2010,Khurgin:2007}.

The average power dissipation $P_d$ and energy relaxation time $\tau_E$ are two key parameters for describing electron energy relaxation in semiconductors under an external electric field. Apparently both quantities depend on the {\it electron temperature}, and
knowing how they depend on $T_e$ is fundamental to the optimization of the HFET devices \cite{Morkoc}.
In experimental studies of hot-electron energy relaxation in GaN-based heterostructures \cite{Matulionis:2005,Matulionis:2008}, the dependence of the electron temperature $T_e$ on the supplied power, which is equal to the power dissipated to the lattice by the hot electrons under steady-state conditions, was {\it directly} measured using the microwave noise technique, and then the electron energy relaxation time as a function of the electron temperature was deduced by $\tau_E(T_e)=k_BdT_e/dPs$, where $P_s$ is the average power supplied to each electron. For the lattice-matched heterostructures the relaxation time was found to fall sharply at the low electron temperatures and decrease very slowly at the high electron temperatures ($>$1200 K) \cite{Matulionis:2008}.

Microscopically, energy relaxation of the hot electrons is governed by scattering with polar-optical phonons. In bulk GaN the polar modes are treated as the longitudinal optical (LO) modes of a single frequency $\omega_{LO}$.  In a simplification therefore the polar-optical phonons of the heterostructures are usually taken to be the LO phonons of the bulk material such as GaN. This is referred to as the three-dimensional phonon (3DP) model. According to the dielectric continuum (DC) model, however the eigenmodes of the polar-optical vibrations in the heterostructure include half-space LO modes and interface modes, and all these modes interact with the electrons in the quasi-2D channel. Both phonon models have been used to study energy relaxation and momentum relaxation for GaAs-based quantum wells \cite{Shah,Ridleyeps}. For GaN-based quasi-2D systems, the 3DP model was used for an early study on the momentum relaxation and low-field electron transport in GaN quantum wells \cite{Anderson}.
In principle, the phonon
eigenmodes of quasi-2D GaN heterostructures should be considered for electron transport studies. Indeed, the DC model has been employed in recent years to study electron-phonon scattering associated with the various phonon modes \cite{Dyson:2011}, as well as the electron momentum relaxation \cite{Zhang:2011} and energy relaxation \cite{Zhang:2013} in GaN heterostructures.
Mori and Ando derived a sum rule \cite{Mori} from the DC model which showed that the sum of the form factors associated with half-space and interface modes was equal to the form factor for bulk phonons. Later, using a microscopic model Register derived a similar sum rule \cite{Register} to prove that the {\it total} electron-phonon scattering rate in the heterostructure was independent of  the phonon basis sets, as long as each set was orthonormal and complete. These sum rules seem to imply that the DC and 3DP phonon models would yield equivalent results for the energy relaxation in GaN heterostructures.
However this claim is only partly true, and caution should be taken in applying the sum rules to the energy relaxation in GaN heterostructures.
First, in Register's scattering rate sum rule polar scattering with electrons is assumed to be made by phonons of a uniform frequency $\omega_{LO}$ \cite{Register} (i.e., all the phonon modes are {\it degenerate}). As is discussed in Appendix A, there is no scattering rate sum rule as the normal modes of the heterostructure including the interface and half-space modes are not degenerate. In fact the interface phonons differ significantly from the half-space phonons in both phonon frequency and electron-phonon coupling strength. Second, as the scalar potential of the interface modes decreases
exponentially from the interface according to $e^{-q|z|}$,
scattering with interface modes is weak in wide wells making the 3DP evaluation accurate. Indeed, for quantum wells with widths greater than 100 $\AA$ the 3DP model suffices for the evaluation of scattering rates \cite{Sarma85,Rucker,Ridleyeps}, energy loss rates \cite{Zhang:1999} and momentum relaxation rates \cite{Zhang:2011}. For narrow wells however scattering with interface phonons becomes increasingly important, and as such, the two phonon models yield quite different rates for GaAs quantum wells \cite{Rucker,Zhang:1999} as well as GaN heterostructures \cite{Zhang:2011}.  For GaN heterostructures with a 30 $\AA$-wide  channel the 3DP model underestimates momentum relaxation rates just below the bulk LO phonon energy by 70\%, and overestimates rates immediately above the LO phonon energy by 40\% compared to the DC model \cite{Zhang:2011}.
However as far as we know there has been no comparison of energy relaxation rates for GaN heterostructures based on the two phonon models. Third, in GaN heterostructures screening from the mobile quasi-2D electrons is strong due to the high electron density. The scattering rate sum rule becomes invalid causing quite different DC and 3DP rates when screening is accounted for. Fourth, the scattering rate sum rule is valid with an equilibrium phonon distribution being taken as an important prerequisite \cite{Nash:1992}. In the GaN HFET where the hot-phonon effect must be taken into account, phonon modes are clearly in  nonequilibrium, with the consequence that different modes make different contributions to the energy relaxation process. In this circumstance it is necessary to use the correct normal modes of the GaN heterostructure to calculate the energy relaxation. Previous calculations for GaAs quantum wells have shown that the energy relaxation rates in the hot-phonon regime depend on the phonon models used \cite{Sarma:1990,Zhang:1999}.

Recently, using the DC model the authors calculated hot-electron energy relaxation in a typical lattice-matched InAlN/AlN/GaN heterostructure \cite{Zhang:2013}. We found that the experimentally observed dramatic fall at low $T_e$ \cite{Matulionis:2008}  was caused chiefly by the fast decreased HPE and electron screening, while the very slow decrease at the high temperatures was due to the fast optical-phonon decay.
In this paper we study the energy relaxation of the hot electrons
in the heterostructure with both DC and 3DP models.
As the 3DP approach is relatively simple and thus convenient
for practical calculation, one of course wants to know how energy relaxation results estimated by this model differ from those calculated with the DC model, and further what causes the discrepancies.
We pay particular attention to a quantitative comparison of the $T_e$-dependencies of the power dissipation and relaxation time
at high temperatures calculated with the two models, as the high $T_e$ relaxation process is of great interest in terms of HFET devices.
This comparison can be made with regard to only the {\it total} power loss and relaxation time. On the other hand, there is an advantage of the use of the DC model, in that the contributions from the quasi-2D phonon modes to the energy relaxation can be singled out.
It is therefore of great interest to find and understand the behaviours of
the half-space modes and interface
modes, in particular those of the {\it interface modes}, in the energy relaxation process in such {\it narrow} channel GaN heterostructures.
The comparison with the 3DP calculations also provides a simple means to examine how the interface phonons contribute.
Equally important from the electron gas side is screening. Clearly dynamic effects of screening from the electrons need to be included owing to the high frequencies of the polar modes.
For these purposes, a comprehensive study needs to be carried out for the energy relaxation in the GaN heterostructures, in which an emphasis is put on how energy relaxation results from the two phonon models differ when both hot phonons and screening are taken into account.
Therefore, using the two phonon models, the power loss and energy relaxation time are calculated as functions of the electron temperature, for a number of phonon scattering processes with electron screening included or excluded.
The energy relaxation results from scattering with the half-space and interface phonons are compared and examined, and then the
{\it total} power loss and relaxation time are further compared with those obtained from the 3DP approximation.
The sum rules for the electron-phonon interactions are closely checked for the {\it pure} phonon emission process (i.e., with no HPE).
For the {\it net} phonon emission process where the hot-phonon reabsorption is accounted for, special attention is paid to the difference in the DC and 3DP calculations to examine the 3DP model in the evaluation of the energy relaxation in particular at high electron temperatures.
We found that the two models yield very close power loss values and relaxation times, the discrepancies being caused by how much the high energy interface phonons contribute to the energy relaxation. With no hot phonons or screening the power loss calculated from the 3DP model is 5\% smaller than the DC power dissipation, whereas slightly larger 3DP power loss (by less than 4\% with a phonon lifetime from 0.1 to 1 ps) is obtained throughout the electron temperature range after including both the HPE and screening.

This paper is organized as follows. In Section II, the DC and 3DP models for lattice-matched InAlN/AlN/GaN heterostructures are  described where the polar phonon modes and associated electron-phonon interactions are given. Then in Section III a formulation of the power dissipation and energy relaxation time in such heterostructures is presented, taking into
account non-equilibrium polar optical phonons, electron degeneracy, and screening from the
mobile electrons. Effective numerical techniques in calculating the generation rates and power loss are also described, in terms of handling the integrals involved.  In Section IV, first we show results of the non-equilibrium phonon
 occupation numbers for both half-space and interface modes in a typical lattice-matched InAlN/AlN/GaN heterostructure.
 These results are used to
analyze how hot phonons slow down the quasi-2D electron energy relaxation in
the high-temperature region. Then, by choosing two GaN heterostructures with different channel widths we compare power dissipation results from the DC and 3DP models for the simple case with no screening. This is to check the sum rules as well as investigate phonon confinement effects and roles the half-space and interface modes play in the respective pure and net phonon emission processes.
In order to examine the usual 3DP approximation in the evaluation of energy relaxation,
we further compare the DC and 3DP results of power loss and energy relaxation time in the lattice-matched heterostructure for a number of detailed phonon scattering processes with or without electron screening.
Comparisons with the experimental data as well as the bulk GaN situation are also made, and the hot-phonon and screening effects are discussed in great detail.
Finally, Section V summarizes the main results obtained.
In Appendix A, starting with the detailed generation rate expressions for the half-space, interface and bulk LO phonons, a restoration of the Mori-Ando sum rule is made. Then the form factor sum rule is used to prove that the power dissipation values obtained from the DC and 3DP models are in general different in the pure emission process, except for the limiting case when scattering with interface phonons is sufficiently weak such as in wide GaN channels, or when the degenerate modes approximation is  imposed as in the study by Register \cite{Register}. These are used to analyze and interpret our energy relaxation results.
In Appendix B, we show the minimum electron kinetic energy for phonon absorption and Fermi-Dirac integrals involved in our energy relaxation calculation, both being functions of the phonon wavevector, which are used for analyzing the non-equilibrium phonon distribution.

\section{Electron-phonon interactions in the DC and 3DP models}

There are two types of polar optical modes in a bulk semiconductor of the wurtzite structure such as GaN and AlN owing to anisotropy. The anisotropy however is very small \cite{Lee}, allowing the semiconductor to be treated as a cubic crystal with LO polar vibration modes \cite{Khurgin:2007,Dyson:2013}. 
The phonon modes of the considered heterostructure In$_x$Al$_{1-x}$N/AlN/GaN is dealt with in a simpler way. As the Indium content in the outer barrier is small ($x<0.2$) the binary alloy In$_x$Al$_{1-x}$N is treated as AlN, the same material as the central barrier. This in effect simplifies the lattice matched
heterostructure In$_x$Al$_{1-x}$N/AlN/GaN as a single heterostructure AlN/GaN \cite{Dyson:2011,Zhang:2013}.
In the dielectric continuum model, the polar vibration modes of a single heterostructure consist of half-space modes and interface modes \cite{Mori}. The half-space modes have the frequencies of the bulk polar modes of the two constituent materials, whose scalar potentials and electric fields occur in the respective constituent regions. The interface modes have different frequencies from the bulk polar modes, and an interface mode has lattice vibrations and electric fields in both constituent regions.

For the In$_x$Al$_{1-x}$N/AlN/GaN heterostructure, let the growth direction be $z$ and the interface between AlN and GaN be at $z=0$, with the barriers in the space
$-L_1<z<0$ (where $L_1=N_1a$, $a$ being the lattice constant) and the electron-containing active region in the space $0<z<L_2$ ($L_2=N_2a$).   Let  $\boldsymbol{\rho}=(x,y)$ be the position vector in the plane parallel to the interface.
When the electrons are completely confined in the GaN channel the half-space modes only in the GaN region interact with the electrons. Now
 the half-space modes in the GaN  region can be indexed by ($q_z$,$\mathbf{q}$), all having the LO frequency $\omega_{LO}$ of bulk GaN.
Here $\mathbf{q}$ is the in-plane phonon wave-vector, and $q_z$ is determined by the fixed end boundary condition imposed on the potential function ($\propto \sin q_zz$), $q_z=l\pi/L_2$ ($l$=1,2,...,$N_2$-1).
The Hamiltonian of an electron interacting with these half-space modes
in the active region 
can be written as
\begin{equation}
\mathcal{H}_h=\sum_{\mathbf{q},q_z}
\frac{\gamma_{LO}}{\sqrt{2V_2}}
\left(\frac{1}{q^2+q_z^2}\right)^{1/2}e^{i\mathbf{q}\cdot\boldsymbol{\rho}}~2\sin q_zz \left[a_{q_z}(\mathbf{q})+a^+_{q_z}(-\mathbf{q})\right],
\label{hami}
\end{equation}
where
$a_{q_z} (\mathbf{q})$ and $a^+_{q_z} (\mathbf{q})$ are the annihilation
and creation operators for the half-space mode of ($q_z$,$\mathbf{q}$). 
$V_2$ is the volume of the GaN half-space $V_2=AL_2$, with $A$ being the sample area, $\gamma_{LO}$ is a constant characterizing the electron-LO-phonon coupling strength, and given by
$\gamma_{LO}^2=2\pi e^2\hbar\omega_{LO}/\epsilon_{LO}$, where $e$ is the electron charge, and  $\frac{1}{\epsilon_{LO}}=(\frac{1}{\epsilon_{\infty}}-\frac{1}{\epsilon_0})$,
$\epsilon_0$ and $\epsilon_{\infty}$ being the static and high-frequency dielectric
constants of bulk GaN.
 We use cgs units throughout the paper.

 The lattice dielectric function of the active GaN material is  $\epsilon(\omega)=\epsilon_{\infty}(\omega^2-\omega^2_{LO})/(\omega^2-\omega^2_{TO})$,
where
$\omega_{TO}$ is the transverse optical (TO) phonon frequency of bulk GaN.
The lattice dielectric function of the simplified barrier AlN is given by
$\bar{\epsilon}(\omega)=\bar{\epsilon}_{\infty}(\omega^2-\bar{\omega}^2_{LO})/(\omega^2-\bar{\omega}^2_{TO})$,
 where $\bar{\epsilon}_{\infty}$, $\bar{\omega}_{LO}$,
$\bar{\omega}_{TO}$ are the high-frequency dielectric constant, the LO and TO phonon frequencies of bulk AlN, respectively. Then
the frequencies of the interface phonons are determined by $\epsilon(\omega)+\bar{\epsilon}(\omega)=0$ \cite{Mori},
which yields two solutions $\omega_n$ ($n$=1,2;
 let $\omega_1 <\omega_2$).
This shows that, similar to the half-space phonons the interface phonons have no dispersion; that is, the phonon frequencies do not depend on the phonon wavevector.
 The interface phonon modes can be simply indexed by ($n$,$\mathbf{q}$) and
the electron-interface-phonon interaction Hamiltonian can be written as
\begin{equation}
\mathcal{H}_i=\sum_{n,\mathbf{q}}\frac{\gamma_n}{\sqrt{2A}}
\frac{1}{\sqrt{q}}~e^{i\mathbf{q}\cdot\boldsymbol{\rho}}~e^{-q\lvert z\rvert}~\left[a_n (\mathbf{q})+a^+_n (-\mathbf{q})\right]
.
\label{hamii}
\end{equation}
Here $\gamma_n$ is the electron-interface-phonon coupling strength,
$\gamma_n^2=2\pi e^2\hbar\omega_n/\epsilon_n$, where $\epsilon_n$ is given by $\frac{1}{\epsilon_n}=2/[\beta^{-1}(\omega_n)+\bar{\beta}^{-1}(\omega_n)]$,  with
$\beta(\omega)$ and $\bar{\beta}(\omega)$ being two dimensionless quantities (thus $\epsilon_n$ is dimensionless),
$\beta(\omega)=\frac{1}{\epsilon_{\infty}}\frac{(\omega^2-\omega_{TO}^2)^2}{\omega^2(\omega_{LO}^2-\omega_{TO}^2)}$,
$\bar{\beta}(\omega)=\frac{1}{\bar{\epsilon}_{\infty}}\frac{(\omega^2-\bar{\omega}_{TO}^2)^2}{\omega^2(\bar{\omega}_{LO}^2-\bar{\omega}_{TO}^2)}$.
$a_n(\mathbf{q})$ and $a^+_n(\mathbf{q})$ are the annihilation
and creation operators for the interface mode ($n$,$\mathbf{q}$).

In the three-dimensional phonon model, the phonon modes are simply bulk LO modes, which are normalized to the sample volume of the entire heterostructure, $V=AL$, with $L=L_1+L_2$ being the dimension of the heterostructure in the growth direction $z$. The bulk LO modes are indexed by the three-dimensional phonon wavevector $\mathbf{Q}=(\mathbf{q},Q_z)$, where the wavevector $Q_z$  is given by the usual periodic boundary condition, $Q_z=n2\pi/L$, $n$ being an integer, $-(N_1+N_2)/2\leq n<(N_1+N_2)/2$. The electron-LO-phonon interaction is given by the Fr\"ohlich interaction
\begin{equation}
\mathcal{H}_{LO}=\sum_{\mathbf{Q}}
\frac{\gamma_{LO}}{\sqrt{V}}
\frac{1}{Q}e^{i\mathbf{Q}\cdot\mathbf{r}} \left[a(\mathbf{Q})+a^+(-\mathbf{Q})\right],
\label{hamiLO}
\end{equation}
where $a(\mathbf{Q})$ and $a^+(\mathbf{Q})$ are the annihilation
and creation operators of the bulk LO mode $\mathbf{Q}$. 
We note that in the growth direction the wavevector $Q_z$ differs from the $q_z$ of the half-space modes. The two sets of discrete wavevector values result from the two different types of boundary conditions on the respective mode potential functions \cite{Ridleyeps}. The bulk phonon modes are planewaves with electric potential $\propto e^{iQ_z z}$ whereas the half-space modes are standing waves with potential function  $\propto \sin q_zz$.

\section{Hot-electron power dissipation and energy relaxation time}

Confinement in the growth direction $z$ quantizes the motion of an electron, and for the GaN-based heterostructures, the narrow and shallow confinement allows us to consider only the lowest subband \cite{Dyson:2011} which is densely populated by the electrons.
Let $\phi(z)$ be the confinement envelope function  corresponding to energy $\epsilon_g$ for the lowest electron subband.
Then the electron wave-function and energy can be written as $\psi_{\mathbf{k}}(\mathbf{r})=\frac{1}{\sqrt{A}}\phi(z)e^{i\mathbf{k}\cdot\boldsymbol{\rho}}$, and $E_{\mathbf{k}}=\epsilon_g+\varepsilon_{\mathbf{k}}$, respectively,  where
$\mathbf{k}$ is the in-plane electron wavevector, and  $\varepsilon_{\mathbf{k}}$ is the electron kinetic energy,
$\varepsilon_{\mathbf{k}}=\hbar^2k^2/(2m^*)$, with
$m^*$ being the electron effective mass.

The degenerate statistics of a high density of electrons is described by the Fermi-Dirac distribution function, $f(E)=1/(1+e^{(E-E_F)/k_BT_e})$, where $E_F$ is the Fermi energy of the quasi-2D electron gas, which is determined by the areal electron density $n_A$ and temperature $T_e$. The thermal equilibrium population of the phonons of frequency $\omega$ at temperature $T$ is given by the Bose-Einstein distribution function, $N(\omega,T)=1/(e^{\hbar\omega/k_BT}-1)$. Knowing the electron-phonon interaction Hamiltonians (Sec. II) the energy relaxation of the hot electrons by scattering with interface and half-space phonons or bulk LO phonons can be calculated by Fermi's golden rule. For a given interface mode ($n$,$\mathbf{q}$) the  {\it number} of phonons which are generated by the hot electrons per unit time can be written as
\[
W_n(\mathbf{q})=\frac{2\pi}{\hbar}\sum_{\mathbf{k}}~|M_{n,\mathbf{q}}(\mathbf{k};\mathbf{k}+\mathbf{q})|^2
\left\{(g_n(\mathbf{q})+1)f(E_{\mathbf{k}+\mathbf{q}})[1-f(E_{\mathbf{k}})]-g_n(\mathbf{q})f(E_{\mathbf{k}})[1-f(E_{\mathbf{k}+\mathbf{q}})]\right\}
\]
\begin{equation}
\times\delta(E_{\mathbf{k}+\mathbf{q}}-E_{\mathbf{k}}-\hbar\omega_n),
\label{wqif0}
\end{equation}
where $M_{n,\mathbf{q}}(\mathbf{k};\mathbf{k}+\mathbf{q})=\langle
\psi_{\mathbf{k}}\vert \mathcal{H}_i \vert \psi_{\mathbf{k}+\mathbf{q}}\rangle$
is the interaction matrix element associated with
electron states $\mathbf{k}$ and $\mathbf{k}+\mathbf{q}$  and phonon mode ($n$,$\mathbf{q}$)
due to the interface phonon scattering, and $g_n(\mathbf{q})$ is the non-equilibrium interface phonon occupation number. For clearness the index for electron spin has been absorbed into the electron wavevectors.

When the two identities
\begin{equation}
 f(E+\hbar\omega)(1-f(E))=[f(E)-f(E+\hbar\omega)]N(\omega,T_e),
\end{equation}
\begin{equation}
 f(E)[1-f(E+\hbar\omega)]=[f(E)-f(E+\hbar\omega)](N(\omega,T_e)+1),
\end{equation}
are used, one finds that Eq.~(\ref{wqif0}) can be transformed into a concise expression
\begin{equation}
W_n(\mathbf{q})=[N(\omega_n,T_e)-g_n(\mathbf{q})]/\tau_n(\mathbf{q}),
\label{genif}
\end{equation}
where  $1/\tau_n(\mathbf{q})$ is referred to as the phonon generation rate
\cite{Khurgin:2007} and given by
\begin{equation}
\frac{1}{\tau_n(\mathbf{q})}=\frac{2\pi}{\hbar}\sum_{\mathbf{k}}~|M_{n,\mathbf{q}}(\mathbf{k};\mathbf{k}+\mathbf{q})|^2
\left[f(E_{\mathbf{k}})-f(E_{\mathbf{k}}+\hbar\omega_n)\right]
\delta(E_{\mathbf{k}+\mathbf{q}}-E_{\mathbf{k}}-\hbar\omega_n).
\label{nuif}
\end{equation}
Expression (\ref{genif}) has a clear physical meaning; that is, $W_n(\mathbf{q})$ is simply the phonon generation {\it number} $\Delta N=N(\omega_n,T_e)-g_n(\mathbf{q})$ divided by the phonon generation {\it time} $\tau_n(\mathbf{q})$ for any particular interface phonon mode ($n$,$\mathbf{q}$).

The polar optical modes decay via the lattice anharmonicity, and the number of interface phonons that decay per unit time is
\begin{equation}
D_n(\mathbf{q})=[g_n(\mathbf{q})-N(\omega_n,T_0)]/\tau_p,
\label{eqstaif}
\end{equation}
where $\tau_p$ is the phonon lifetime which is assumed to have the same value for all polar modes \cite{Sarma:1990,Ridleyeps}, 
 and $N(\omega_n,T_0)$ is the thermodynamic equilibrium interface phonon number at the lattice temperature $T_0$.
At steady state then one has $W_n(\mathbf{q})=D_n(\mathbf{q})$ \cite{Zhang:2013,Dyson:2012}, from which one finds the nonequilibrium interface phonon occupation number
\begin{equation}
g_n(\mathbf{q})=\frac{\frac{1}{\tau_n(\mathbf{q})}N(\omega_n,T_e)+\frac{1}{\tau_p}N(\omega_n,T_0)}{\frac{1}{\tau_n(\mathbf{q})}+\frac{1}{\tau_p}}.
\label{gnqif}
\end{equation}
This shows that the hot phonon occupation number depends on the relative magnitude of the phonon generation and decay rates $1/\tau_n(\mathbf{q})$ and $1/\tau_p$, with $g_n(\mathbf{q})\approx N(\omega_n,T_e)$ when $1/\tau_n(\mathbf{q})\gg 1/\tau_p$, and $g_n(\mathbf{q})\approx N(\omega_n,T_0)$ when $1/\tau_n(\mathbf{q})\ll 1/\tau_p$, the latter case being equivalent to neglecting the hot-phonon effect.  As $\lim_{q\rightarrow 0}\frac{1}{\tau_n(\mathbf{q})}=\lim_{q\rightarrow \infty}\frac{1}{\tau_n(\mathbf{q})}=0$ (refer to Appendix B), from Eq.~(\ref{gnqif}) then one readily finds $\lim_{q\rightarrow 0}g_n(\mathbf{q})=\lim_{q\rightarrow \infty}g_n(\mathbf{q})=N(\omega_n,T_0)$.

Similarly, the  {\it number} of half-space phonons generated by the hot electron gas per unit time can be expressed as
\begin{equation}
W_{q_z}(\mathbf{q})=[N(\omega_{LO},T_e)-g_{q_z}(\mathbf{q})]/\tau_{q_z}(\mathbf{q}),
\label{genhs}
\end{equation}
where the generation rate is
\begin{equation}
\frac{1}{\tau_{q_z}(\mathbf{q})}=\frac{2\pi}{\hbar}\sum_{\mathbf{k}}~|M_{{q_z},\mathbf{q}}(\mathbf{k};\mathbf{k}+\mathbf{q})|^2
\left[f(E_{\mathbf{k}})-f(E_{\mathbf{k}}+\hbar\omega_{LO})\right]
\delta(E_{\mathbf{k}+\mathbf{q}}-E_{\mathbf{k}}-\hbar\omega_{LO}).
\label{nuhs}
\end{equation}
Here $M_{q_z,\mathbf{q}}(\mathbf{k};\mathbf{k}+\mathbf{q})=
\langle
\psi_{\mathbf{k}}\vert \mathcal{H}_h \vert \psi_{\mathbf{k}+\mathbf{q}}\rangle$
is the interaction matrix element associated with
electron states $\mathbf{k}$, $\mathbf{k}+\mathbf{q}$ and phonon mode ($q_z$,$\mathbf{q}$)
due to the half-space phonon scattering, $g_{q_z}(\mathbf{q})$ is the hot  half-space phonon occupation number, which is given by Eq.~(\ref{gnqif})
with $\omega_n$ being replaced with $\omega_{LO}$ and $1/\tau_n(\mathbf{q})$  replaced by the half-space phonon generation rate $1/\tau_{q_z}(\mathbf{q})$.

Therefore, in the DC model the average power dissipated per electron is given by
\begin{equation}
P_d=\frac{1}{n_AA}\sum_{\mathbf{q}}\left\{\sum_n\hbar\omega_n[N(\omega_n,T_e)-g_n(\mathbf{q})]/\tau_n(\mathbf{q}) +\sum_{q_z>0}\hbar\omega_{LO}[N(\omega_{LO},T_e)-g_{q_z}(\mathbf{q})]/\tau_{q_z}(\mathbf{q})\right\},
\label{pdDC}
\end{equation}
the right hand side being a sum of the contributions from both the interface and half-space phonons.

In the 3DP model the average power dissipation per electron is simply given by
\begin{equation}
P_d=\frac{\hbar\omega_{LO}}{n_AA}\sum_{\mathbf{Q}}\left[N(\omega_{LO},T_e)-g(\mathbf{Q})\right]/\tau(\mathbf{Q}),  
\label{pd3DP}
\end{equation}
where the generation rate $1/\tau(\mathbf{Q})$ has the same expression as $1/\tau_{q_z}(\mathbf{q})$ above [Eq.~(\ref{nuhs})] except that the Fr\"ohlich interaction matrix element should be used instead, and
$g(\mathbf{Q})$ is the hot bulk LO phonon occupation number, whose expression is given by Eq.~(\ref{gnqif}) after substituting the rate $1/\tau(\mathbf{Q})$ for $1/\tau_n(\mathbf{q})$ and the phonon frequency $\omega_{LO}$ for $\omega_n$.  

Knowing the power dissipation $P_d$ then a hot-electron energy relaxation time $\tau_E$ can be defined in the hydrodynamic model \cite{Dyson:2012}  through
\begin{equation}
 P_d(T_e)=k_B(T_e-T_0)/\tau_E. 
\end{equation}

In-plane isotropy is used to simplify the energy relaxation calculations.
Then the in-plane phonon wave-vector dependent quantities such as the phonon generation times $\tau_n(\mathbf{q})$,  $\tau_{q_z}(\mathbf{q})$,  $\tau(\mathbf{Q})$ and phonon occupation numbers $g_n(\mathbf{q})$, $g_{q_z}(\mathbf{q})$, $g(\mathbf{Q})$ are reduced to functions of only the magnitude $q$. Therefore the summation over wavevector $\mathbf{q}$
 [Eqs.~(\ref{pdDC}) and (\ref{pd3DP})] is converted to a double integral (over $q$, $\theta$) and then reduces to a single integral over only $q$. We also note that, as neither the half-space, interface phonons nor bulk LO phonons have dispersion (i.e., the phonon frequencies do not depend on wavevector $\mathbf{q}$), the $q$-dependence of the hot phonon occupation number is dictated by the variation of the generation rate with $q$ [refer to Eq.~(\ref{gnqif}), for instance for the interface modes].

We now include screening.
The electron-phonon interactions are
screened by the mobile electrons.
The response of the electron plasma to a polar disturbance  from the lattice is encapsulated by the dielectric function of the electron gas. For a high temperature electron gas as considered here,  the Boltzmann distribution function is used to approach the energy distribution of the hot electrons. Then the Lindhard dielectric function of the quasi-2D electron gas reduces to the following form
\begin{equation}
\epsilon(q,\omega)=1+F(q)\frac{\kappa_D}{q}\frac{1}{2a}[Z(y-\frac{1}{2}a)-Z(y+\frac{1}{2}a)],
\label{dfzxz}
\end{equation}
where $Z(s)$ is the plasma dispersion function \cite{Fried}
\begin{equation}
Z(s)=\frac{1}{\sqrt{\pi}}\int_{-\infty}^{\infty}\frac{e^{-x^2}}{x-s}dx,
\label{Zint}
\end{equation}
with $s$ being complex, and
$\kappa_D$ is the two-dimensional Debye screening wavenumber, $\kappa_D=2\pi n_A e^2/(\epsilon_0k_BT_e)$. In the $\epsilon(q,\omega)$ expression  the real arguments of the plasma dispersion function are determined by the two dimensionless quantities $y$ and $a$,
\begin{equation}
y=\left(\frac{m^*}{2k_BT_e}\right)^{\frac{1}{2}}\frac{\omega}{q}, ~~~a=\left(\frac{\hbar^2q^2}{2m^*k_BT_e}\right)^{1/2}.
\end{equation}
In Eq.~(\ref{dfzxz}) $F(q)$ is a form factor which accounts for the confinement effect on the electron-electron Coulomb interaction due to the {\it finite} effective channel width of the heterostructure \cite{Ridleyeps},
\begin{equation}
F(q)=\int dz\int dz'\lvert\phi(z)\rvert^2 \lvert\phi(z')\rvert^2 e^{-q\lvert z-z'\rvert},
\label{ffactorB}
\end{equation}
where $\phi(z)$ is confinement envelope function of the the lowest subband. This form factor is equal to the form factor which was  introduced to describe the electron-bulk-LO-phonon interaction \cite{Mori,Ridleyeps}, namely $F_B(q)$ given by Eq.~(\ref{formFB}).

To simplify calculations the plasma dispersion function [integral expression (\ref{Zint})] is  approached by a two-pole pad\'{e} approximant \cite{Lowe},
\begin{equation}
Z(s)=\frac{i\sqrt{\pi}+(\pi-2)s}{1-i\sqrt{\pi}s-(\pi-2)s^2}.
\label{Zpad}
\end{equation}

Using the properties of the plasma dispersion function \cite{Fried}, it is found that in the static limit  $\omega\rightarrow 0$  the dielectric function Eq.~(\ref{dfzxz}) reduces to
\begin{equation}
\epsilon(q,0)=1+F(q)\frac{\kappa_D}{q}.
\label{epslnq0}
\end{equation}
This is the familiar Debye screening formula which is used to evaluate screening for high-temperature non-degenerate electron gases \cite{Ridleyeps}.
In this study, screening from the mobile electrons is handled by dividing the
 scattering potential, or equally the electron-phonon
interaction
matrix elements by the dielectric function of the quasi-2D electron gas.
The polar disturbance is of course not static and occurs at the {\it finite}
frequency of a particular phonon mode, for instance, a half-space or interface mode.  To account for the
 dynamic effect of screening the frequency $\omega$ in the dielectric function is substituted for the corresponding phonon frequency, and the dielectric function is treated as a function of wave-vector $q$ for each finite phonon frequency.

The  phonon generation rates $1/\tau_n(q)$, $1/\tau_{q_z}(q)$,  $1/\tau(\mathbf{Q})$ are key quantities in the calculation of the power loss and energy relaxation time. The delta function in the rate expressions [Eq.~(\ref{nuif}), for instance]  reflects energy conservation in the scattering of an electron with a phonon of particular mode.
The summation over electron wavevector $\mathbf{k}$ is converted to a double integral,  which is reduced to a form that is proportional to the difference of two complete Fermi-Dirac integrals of order -1/2 [see Appendix A, Eq.~(\ref{nunqa1}) for $1/\tau_n(q)$].
Accurate evaluation of the integral $F_{-\frac{1}{2}}(x)$ [Eq.~(\ref{fdm12})] is important in obtaining the correct energy relaxation results.
The calculation should also be efficient as the integration values are input to calculating the generation rates for all interface and half-space or  bulk modes in a large phonon wavevector space.
To calculate the Fermi-Dirac integral  the integrand  is transformed to $e^{-x}\frac{e^{\mu}\sqrt{x}}{(1+e^{\mu-x})^2}$ such that the Gauss-Laguerre quadrature technique is used to achieve fast and excellent convergence (25 quadrature points are used).  Then these generation rates are inserted into Eqs.~(\ref{pdDC}) and (\ref{pd3DP}) to calculate power dissipation $P_d$. Again the summation over phonon wavevector $\mathbf{q}$ is transformed to a double integral that is reduced to an integral over only $q$ by in-plane isotropy. Numerical integration is then carried by using the Gauss-Legendre quadrature method, 105 quadrature points being used with the cut-off of  $q$ taken to be 8$k_0$ ($k_0$ is the threshold electron wavevector for  LO phonon emission, $k_0=\sqrt{2m^*\omega_{LO}/\hbar}$).
To calculate energy relaxation one also needs the Fermi energy $E_F$, which is determined by the equation $n_AA=\sum_{\mathbf{k}}f(E_{\mathbf{k}})$.  Therefore one finds that the Fermi energy is given by $E_F=\epsilon_g+k_BT_e\ln(e^{n_A/n_{T_e}}-1)$, where $n_{T_e}$ has the dimension of areal density, $n_{T_e}=m^*k_BT_e/(\pi\hbar^2)$.

We model the electron envelope function $\phi(z)$ for the triangular potential well by the Fang-Howard wave-function \cite{Fang,Stern},
\begin{equation}
\phi(z)=\sqrt{\frac{b^3}{2}}ze^{-bz/2},
\label{Fangwf}
\end{equation}
where $b$ is a variational parameter which is determined by minimizing the total energy of the quasi-2D electron gas. $b$ is related to the areal electron density $n_A$ in the GaN channel via
\begin{equation}
b=\left(\frac{33\pi e^2m^*n_A}{2\epsilon_0\hbar^2}\right)^{1/3}.
\label{bpara}
\end{equation}
In this wavefunction model
an effective channel width $d$ is defined as twice the average penetration depth of the charge in the active GaN region \cite{Ando,Anderson01}; $d$
is related to the Fang-Howard $b$ parameter via $d=6/b$.

In this study, the material parameters are taken from Refs. \cite{Nipko,Bulutay,Bulutay02}. The LO and TO phonon frequencies used for GaN are $\omega_{LO}$=91.13 meV,  $\omega_{TO}$=66.08 meV, and for AlN we use  $\omega_{LO}$=110.7 meV, $\omega_{TO}$=76.1 meV. The high-frequency dielectric constants are taken to be 5.29 and 4.68 for bulk GaN and AlN respectively. The electron effective mass for GaN is $m^*=0.22m_0$ ($m_0$ is the free electron mass),  and the lattice temperature is fixed at room temperature 300 K.
The optical phonon life-time is a key parameter in the electron energy relaxation study. Thus a range of optical phonon life-time values from 0.1 to 2 ps are taken to examine the hot-phonon effect.

\section{Results and discussions}

\subsection{Non-equilibrium phonon occupation number}

The optical phonons contributing to hot-electron energy relaxation are the GaN half-space modes ($\hbar\omega_{LO}$=91.13 meV), and the lower- and higher-energy interface modes ($\hbar\omega_1$=69.70 and $\hbar\omega_2$=102.09 meV,
respectively).
We first look at the hot-phonon occupancy in phonon wavevector space, as a large number of non-equilibrium phonons of these modes are generated during  energy relaxation when their decay is much slower than their emission by the hot electrons.  To do this we choose an electron temperature of 1000 K for the electron gas of the areal density of  $1.2\times 10^{13}$/cm$^2$ in a typical lattice-matched InAlN/AlN/GaN heterostructure. The high-$T_e$ experimental value of optical-phonon lifetime 0.1 ps is used for all half-space and interface modes \cite{Matulionis:2008}.  
We calculated the hot-phonon occupation numbers as  functions of the in-plane phonon wavevector $q$ for the half-space modes, $q_z=l\pi/L_2$, with mode indices
$1 \leq l\leq 100$, as most of these phonons participate in significant electron-phonon scattering. We found that for a given wavevector $q$ the half-space phonon occupation number increases with the mode index $l$ and then decreases after it reaches the maximum value of a certain $l_a$ mode. This is illustrated in Fig.~\ref{fig1}(a) where the wavevector $q$-dependent hot-phonon population are shown for a number of different orders of half-space modes as labelled by the mode indices $l$. This result can be explained as follows. The hot electrons are confined in a very narrow channel with an effective width of only 44 $\AA$, whereas the half-space phonons interacting with the electrons are present in the entire GaN region ($0<z<L_2$). Thus the electron-half-space-phonon overlap integral $\Gamma_H(q_z)$ [Eq.~(\ref{Ihqz})] as well as the squared interaction matrix element [$\propto \Gamma_H^2(q_z)/(q^2+q_z^2)$] strongly depends on $q_z$ or equally the mode index $l$, as displayed in Fig.~\ref{fig1}(b).  According to Eq.~(\ref{nuzqa1}), therefore the $q_z$-dependence of the squared interaction matrix element dictates the phonon generation rate $1/\tau_{q_z}(q)$ and the variation of the half-space phonon population with $q_z$.
This is quite different from what happens in the usual {\it square} quantum wells such as GaAs/AlGaAs quantum wells, where the hot-phonon population decrease as the confined-mode index in the growth direction $z$ increases \cite{Zhang:1999}. This is because both the electrons and phonons are confined in the same {\it well} region \cite{Ridleyeps}, resulting in stronger electron-phonon interaction with a larger overlap integral for the lower-order phonon mode than the higher-order mode.
We also see from Fig.~\ref{fig1}(a) that the half-space phonons
of different orders $l$ have peak occupation numbers occurring at different phonon wavevectors $q$. In the two limits $q\rightarrow 0$ and $q\rightarrow +\infty$, however the occupation numbers of all order half-space modes approach the common thermal equilibrium value at room temperature, $N(\omega_{LO},T_0)=1/(e^{\hbar\omega_{LO}/k_BT_0}-1)=0.03$. 
The interface phonon occupation numbers are shown in Fig.~\ref{fig2}(a) for both the lower-frequency ($\omega_1$) and higher-frequency ($\omega_2$) interface modes. The lower-frequency phonons $IF_1$ have a smaller population, whilst the higher-frequency modes $IF_2$ have a higher peak and are densely populated in a broader $q$-region where the $IF_2$ phonon generation rate $1/\tau_2(\mathbf{q})\gg 1/\tau_p$ because they have much greater electron-phonon coupling strength than the $IF_1$ modes (the Fr{\"o}hlichlike coupling constants for the interface modes, $\alpha_n=\frac{e^2}{\hbar}(\frac{m^*}{2\hbar\omega_n})^{1/2}/\epsilon_n$ \cite{Mori}, are $\alpha_1$=0.02 and $\alpha_2$=0.5; that is, the higher frequency IF$_2$ phonons have more than one order of magnitude larger coupling strength than the lower frequency IF$_1$ phonons). According to Eq.~(\ref{gnqif}), the occupation number $g(q)$ of the frequency $\omega$ modes is restricted to the range $N(\omega,T_0)\leq g(q)\leq
N(\omega,T_e)$. Therefore the peak occupation number of the $\omega_n$ interface phonons is smaller than $N(\omega_n,T_e)$, namely the Bose-Einstein distribution function at electron temperature $T_e$.  Further, 
with $\omega_1<\omega_{LO}<\omega_{2}$, the minimum occupation number of the half-space modes is larger than that of the higher-frequency $IF_2$ modes but smaller than the minimum occupation number of the lower-frequency $IF_1$ modes [comparing Fig.~\ref{fig1}(a) and Fig.~\ref{fig2}(a)].
For both half-space and interface phonons, the occupation number curves [Figs.~\ref{fig1}(a) and \ref{fig2}(a)] have a {\it steep} edge on the small-wavevector side and a slow slope on the large-$q$ side. We  found that this originates from the wavevector $q$-dependence of the energy $\Delta_q$ [that is, the {\it minimum} electron kinetic energy for phonon absorption, Eq.~(\ref{epsqhs})], as is shown in Fig.~\ref{fig9}(a) and discussed in Appendix B. Further, we found that the $q$-dependent non-equilibrium phonon occupation number as shown above is governed by the Fermi-Dirac integral $F_{-\frac{1}{2}}(\xi_q)$ [with $\xi_q$ being given by Eq.~(\ref{xizths})] as a function of the phonon wavevector $q$ [Fig.~\ref{fig9}(b)], which is discussed in Appendix B.

The influence of screening from electrons on hot-phonon population is shown in Fig.~\ref{fig2}(b) for the interface
modes. Compared to the non-screening calculation [Fig.~\ref{fig2}(a)], static Debye screening has significantly reduced the phonon population, and in particular the population of the $IF_1$ phonons
are reduced substantially as these low-frequency phonons are restricted to only a small-$q$ region ($q\ll k_0$) of wavevector space where screening from the electrons is very strong with large $F(q)/q$ and hence large values of dielectric function $\epsilon(q,0)$ [Eq.~(\ref{epslnq0})].
Recall that $F(q)$ is the form factor associated with the Coulomb interaction [Eq.~(\ref{ffactorB})], which increases as the wavevector $q$ decreases.
When dynamic screening is used [Fig.~\ref{fig2}(b), dashed and dotted lines], we see that the interface phonon population becomes narrower in wavevector $q$-space, with the occupation numbers at small $q$ wavevectors being increased rather than decreased, compared to the case of excluding screening in Fig.~\ref{fig2}(a).
This antiscreening arises due to the dispersion of the quasi-2D electron plasma oscillation frequency, namely $\omega_{pl}(q)=\sqrt{2\pi e^2nqF(q)/(\epsilon_0 m^*)}$, which is smaller than the phonon frequency at long wavelengths \cite{Ridleyeps}. In this circumstance the electron plasma cannot move sufficiently fast to cause screening to the polar disturbance from the lattice \cite{Ridleyqps}.
These hot phonon population results will be used to analyze the power dissipation calculation below.

\subsection{Hot electron power dissipation and energy relaxation time}

The three-dimensional phonon (3DP) model has been widely used to evaluate electron energy and momentum relaxation rates for quasi-2D semiconductor systems \cite{Ridley82,Riddoch,Ridleyeps,Stroscio}.
Here we compare hot electron power dissipation in GaN based heterostructures calculated with the DC and 3DP models.
First we examine the sum rules as applied for the energy relaxation in GaN heterostructures.  We include only the hot phonon effect and do not consider  screening. 
To do this, we consider two heterostructures with different {\it effective} channel widths, namely a strained Al$_{0.05}$Ga$_{0.95}$N/GaN heterostructure \cite{Ambacher02} with a wide well of 110 $\AA$ (corresponding to an areal electron density of 7 $\times10^{11}$/cm$^2$) and a lattice matched In$_{0.18}$Al$_{0.82}$N/AlN/GaN heterostructure \cite{Matulionis:2008} with a narrow well of 44 $\AA$ (corresponding to electron density 1.2$\times10^{13}$/cm$^2$).   Then the average
power dissipated per electron is calculated as functions of the electron temperature with hot phonons being excluded or included,
for the latter case two phonon life-time values being used, $\tau_p$=0.5 and 2 ps, to investigate the HPE.
Figs.~\ref{fig3}(a) and \ref{fig3}(b) show the results with the  electron temperatures ranging from room temperature up to 2500 K.
For the strained heterostructure with a wide channel,
as shown in Fig.~\ref{fig3}(a) the two phonon models yield literally the same power dissipation.
In this case, the interface phonon scattering with potential decreasing exponentially according to $e^{-q|z|}$ [Eq.~(\ref{hamii})] is very weak, as the average value of the position for electrons $\bar{z}$ which is half the effective channel width \cite{Mori} is $\bar{z}$=55 $\AA$, whereas the characteristic wavevector $k_0$ for the interface phonons is around 0.07 1/$\AA$, making $k_0\bar{z}\approx$ 4.
The form factor for the interface phonons [Eq.~(\ref{formFI})] $F_I(q)$ can be neglected, and as is proved in Appendix A, then power dissipation values given by the two phonon models are  equal, which is consistent with both sum rules \cite{Mori,Register}. 
For the heterostructure with a narrow 44$\AA$ channel,  in contrast,
the interface phonon scattering is significantly enhanced, and in this case the two phonon models yield different power loss values (see proof in Appendix A). Therefore a clear difference is seen between the power dissipation curves calculated with the two phonon models [Fig.~\ref{fig3}(b)].
In the pure emission case the DC model yields higher power dissipation than the 3DP approximation. Taking the HPE into account, however the 3DP power loss becomes larger. To find the cause we need to separate and check contributions to the power dissipation from the half-space modes, the lower and higher frequency interface modes respectively [Fig.~\ref{fig4}(a)].
We see that the half-space phonons dominate the energy relaxation process (due to the large density of states of the half-space modes)
and contribute larger power dissipation than the interface phonons, whilst the lower-frequency IF$_1$ phonons make only a small contribution due to their low energy, small coupling strength and narrow population distribution in $q$-space (refer to Fig.~\ref{fig2} and the preceding subsection).
Without HPE the power dissipation due to the {\it high-energy} IF$_2$ phonons increases rapidly with $T_e$ in particular above 1000 K compared to the half-space phonons. For instance,  the IF$_2$ power dissipation at $T_e$=2500 K has risen to 46\% the power dissipation due to the half-space phonons.
This causes a larger total power dissipation with the DC model than the 3DP model [Fig.~\ref{fig3}(b)].
When the hot phonons are taken into account,
however we see from Fig.~\ref{fig4}(a) that the power dissipation due to the IF$_2$ phonons drops dramatically by about 85\%,
because the IF$_2$ phonon generation number per unit time $W_2(q)$ [Eq.~(\ref{genif})] is substantially reduced [compare the two curves in  Fig.~\ref{fig4}(b)], as  the nonequilibrium IF$_2$ phonons with a broad population distribution in wavevector space (as illustrated in Fig.~\ref{fig2}) are re-absorbed.
As a result the total DC power dissipation becomes smaller than that evaluated with the 3DP model [Fig.~\ref{fig3}(b)]. Therefore the difference between the DC and 3DP results is due to the IF$_2$ phonons - their contribution to the power loss is enhanced at the high electron temperatures in pure emission but is dramatically reduced after including the HPE.
In recent studies on GaN heterostructures, we found that interface phonon absorption causes negative momentum relaxation rates \cite{Zhang:2011},
and also an increased interaction with the $IF_2$ modes leads to a reduction of phonon lifetime \cite{Dyson:2011}.

In both heterostructures the calculations with both phonon models  [Figs.~\ref{fig3}(a) and \ref{fig3}(b)]  show that the electron power dissipated increases rapidly with $T_e$ at low temperatures but  the increase becomes slower at high temperatures, similar to that which  occurs in Si-doped bulk GaN \cite{Zhang:2014}.
In the simpler case with no screening or hot phonons, the phonon generation number is $\Delta N(T_e)=N(\omega_{LO},T_e)-N(\omega_{LO},T_0)$ for bulk LO modes, and the temperature dependence of the power loss $P_d$, according to Eq.~(\ref{pd3DP}), is determined entirely by the product of the generation number $\Delta N(T_e)$ and the {\it total} generation rate of all bulk modes $\sum_{\mathbf{Q}}\frac{1}{\tau(\mathbf{Q})}$.  The low-temperature power loss is dictated by the {\it exponential} increase of $\Delta N(T_e)$ with $T_e$, $\Delta N(T_e)\approx e^{-\hbar\omega_{LO}/(k_BT_e)}-e^{-\hbar\omega_{LO}/(k_BT_0)}$. At high temperatures ($>$1200 K),
whilst the generation number $\Delta N(T_e)$ increases with $T_e$ the total generation rate is reduced, resulting in  the {\it slow} rise of power dissipation.
Physically the generation rate decreases as the difference in electron occupation numbers within the phonon energy $\hbar\omega_{LO}$ becomes smaller at a higher electron temperature.

We now include screening to make a comprehensive study of the electron power dissipation calculated with the DC and 3DP models. When both screening and hot phonons are included, strictly speaking the scattering rate sum rule is not applicable, and then one needs to find how much discrepancy the 3DP evaluation yields with respect to the DC calculation. Thus calculations were performed using each phonon model for a number of cases, namely, (i) excluding hot phonons and screening, (ii) including only the HPE, (iii) including only static (Debye) screening, (iv) including only dynamic screening, (v) including both HPE and static screening, and (vi) including both the HPE and dynamic screening.
The results are show in Fig.~\ref{fig5} for the lattice-matched In$_{0.18}$Al$_{0.82}$N/AlN/GaN heterostructure (with a 44-$\AA$-wide channel), where a polar optical phonon lifetime of 1 ps is used for all the cases of including the HPE.
Several points can be made by comparing the power dissipation values in the various cases. First, the power dissipation is substantially reduced by static screening [compare cases (i) and (iii)], whereas the reduction using dynamic screening is much smaller,  which is only $\sim$ 30\% the reduction caused by Debye screening [compare cases (i), (iii) and (iv)].
When both screening and hot phonons are included, the  power loss  values obtained with the static and dynamic screening models get closer as $T_e$ increases; at $T_e$=2500 K the power loss is 14\% smaller from Debye screening than from the dynamic screening model.
Second, at the low temperatures interestingly both phonon models yield enhanced rather than slowed power dissipation when {\it dynamic} screening is included. That is, {\it anti-screening} occurs when the electron temperature $T_e$ is lower than 840 K for the DC model or $T_e$$<$770 K for the 3DP model.
This is explained as follows. Expressions (\ref{pdDC}) and (\ref{pd3DP}) show that  mathematically the power dissipation is a sum of the contributions that are connected with the various in-plane phonon wavevectors $\mathbf{q}$.
When dynamic screening is taken into account, the bare electron-phonon interaction is screened or anti-screened depending on wavevector \cite{Ridleyqps}. The large-$q$ components in the summation contribute screening while the small-$q$ components which are connected to the slow motion of the electron gas contribute anti-screening \cite{Ridleyeps}, due to the dispersion of the quasi-2D electron plasma frequency.
At the low electron temperatures, the small-$q$ components dominate as the degenerate distribution of the dense electrons favours the electron-phonon scattering with small transfer wave-vectors $q$. 
At the high temperatures, on the other hand, the electrons are distributed over a large $k$-space, and they cause screening when the large $q$-components dominate. We note that antiscreening for the quasi-2D electron energy relaxation was observed early in GaAs quantum wells and occurred also at low electron temperatures \cite{Sarma:1988}. 
Third, for the three cases with no hot phonons, namely (i), (iii), (iv), the DC model yields higher power dissipation than the 3DP model. However taking into account hot phonons, namely in the corresponding cases (ii), (v), (vi) the
3DP power loss becomes larger. 
What causes this has become clear after our discussion above for Fig.~\ref{fig3}(b); that is,  it is due to the higher-energy interface phonons.  Fourth, throughout the temperature range
the 3DP power loss is 5\% smaller than the DC power dissipation  in the simplest case (i), but becomes larger (by less than 4\%) after including both HPE and screening as in cases (v), (vi). A similar deviation is obtained when reducing the phonon lifetime to 0.1 ps except for the static screening case where only a tiny 0.2\% deviation occurs. The 3DP model yields such a close result  to the DC calculation, because including screening does not alter the order of the DC and 3DP power loss values in terms of their relative magnitude [that is, the DC power dissipation is higher. Refer to cases (i), (iii), (iv)], while accounting for the HPE does alter the order [compare cases (i), (ii)].

Experimentally, using the microwave noise technique the electron temperature $T_e$ as a function of the supplied power $P_s$ was  directly measured
for Si-doped bulk GaN \cite{Liberis:2006}, strained AlGaN/GaN
 \cite{Matulionis:2005} and lattice-matched In$_{0.18}$Al$_{0.82}$N/AlN/GaN  \cite{Matulionis:2008} heterostructures.
The total number of electrons was estimated from the measured low-field Hall mobility and channel resistance using Ohm's law.
Under steady-state conditions, the supplied power is equal to  the {\it total} power dissipated to the lattice by the hot electrons. Then one can obtain the experimental data of the {\it average} power dissipation per electron versus electron temperature (see Fig. 4 of Ref.\cite{Matulionis:2008} for the lattice-matched In$_{0.18}$Al$_{0.82}$N/AlN/GaN heterostructure). The power loss was shown to increase with the electron temperature (from 2 nW/electron at $T_e$=500 K, for instance, to 150 nW/electron at $T_e$=2500 K) but the dependence is complicated by electron screening  and the variation of the polar optical phonon lifetime with $T_e$, as was discussed in our previous study \cite{Zhang:2013}.
In the simple approximation where neither hot phonons nor screening is included, the calculated power dissipation is four times as large as the experimental values. Accounting for screening and HPE brings the calculation much closer to the experimental data. However the theoretical values remain over 2.5 times higher in the low temperature region  even with static screening which is generally believed to overestimate the screening effect, and the use of large phonon lifetimes there such as 20 ps will produce a fit with experiment.

To quantify the screening effect and/or HPE a reduction factor $\beta$ is introduced, $\beta=P_d^0/P_d$, where $P_d^0$ is the power dissipation without screening or hot phonons, and $P_d$ is the corresponding power loss when screening and/or the hot phonons are included.
Fig.~\ref{fig6} shows the temperature-dependences of the reduction factors associated with only the HPE, only screening (Debye screening or dynamic screening), and both HPE and screening calculated with the DC and 3DP phonon models as labeled (using a polar optical phonon lifetime of 1 ps) for the lattice-matched In$_{0.18}$Al$_{0.82}$N/AlN/GaN heterostructure.  When only screening is included almost the same reduction factors are obtained from the two phonon models, with the two curves coinciding for either static or dynamic screening case. We see anti-screening again in the dynamic screening alone case for electron temperatures lower than about 800 K (below the dotted horizontal line $\beta$=1 in Fig.~\ref{fig6}) with reduction factor $\beta<$1, as antiscreening causes faster power dissipation $P_d$ than $P_d^0$, $P_d>P_d^0$.  In all the other cases, as $\beta>1$ the electron energy relaxation has slowed down after including screening and/or hot phonons. We see a stronger HPE at low electron temperatures, with the reduction factors decreasing with increasing the electron temperature.
The 3DP model underestimates the HPE as expected, and as a result the reduction factors from the 3DP calculation are smaller than the DC result even when screening is included. The reduction factor is $\sim$9\% smaller by the 3DP approach than by the DC model in the high-$T_e$ region. Using either of the two screening models the reduction factor associated with both hot phonons and screening decreases as $T_e$ increases, in both the DC and 3DP calculations, but the high-$T_e$ reduction factor tends to be flat and the $\beta$ values from the static and dynamic screening models are quite close, with $\beta$ varying only from 2.5 to 3.2. That is to say, with a phonon lifetime of 1 ps the high-temperature electron power loss is reduced approximately  by a factor of 3  due to screening and the HPE.

We now turn to the energy relaxation time $\tau_E$. Fig.~\ref{fig7}  shows the dependences of the electron energy relaxation times  on the electron temperature for the lattice-matched heterostructure, calculated with the DC and 3DP models for three cases, namely, (i) excluding hot phonons and screening, (ii) including both the HPE and static (Debye) screening, and (iii) including both HPE and dynamic screening.
With no hot phonons or screening, both phonon models yield relaxation times around 0.05 ps but at low electron temperatures $T_e <$ 500 K  there is a drop in relaxation time $\tau_E^0$ ($\tau_E^0$ is the energy relaxation time with no screening or HPE, $\tau_E^0=\frac{k_B(T_e-T_0)}{P_d^0}$),
 which is caused by the exponential rise of the generation number $\Delta N(T_e)$ with $T_e$.
When screening and hot phonons are taken into account, the energy relaxation time can be conveniently expressed as $\tau_E=\frac{k_B(T_e-T_0)}{P_d^0}\frac{P_d^0}{P_d}=\tau_E^0\beta$, where $\beta$ is the reduction factor caused by screening and HPE.
There is a very small difference in the relaxation times calculated from the two phonon models (upper two pair of curves with $\tau_p$=1 ps in Fig.~\ref{fig7}), with the 3DP relaxation times being  4\% smaller than the DC ones at high electron temperatures (Fig.~\ref{fig8}) and the deviation staying within 5\% when reducing the phonon lifetime to 0.1 ps.
As the combined hot phonon and screening effect, parametrized by the reduction factor, decreases as $T_e$ is elevated (refer to Fig.~\ref{fig6} above),
a great fall of energy relaxation time appears at temperatures $T_e<$ 750 K (Fig.~\ref{fig7}). In particular the fall is sharp when Debye screening is used as the screening wavenumber $\kappa_D$ ($\kappa_D\approx 5.8k_0T_0/T_e$) decreases fast with $T_e$ (at $T_e$=1000 K, for instance, $\kappa_D$ reduces to 1.7$k_0$).

At {\it high} temperatures (above 1200 K), on the other hand,
 the relaxation time, $\tau_E$,  stays almost flat with a very small and slow rise when a single phonon lifetime 1 ps is used throughout the temperature range. The relaxation time is 0.15 ps when static screening is used, which is slightly larger than the $\sim$0.12 ps value obtained with dynamic screening. This  saturation in energy relaxation means that the increases in the average electron kinetic energy and power dissipation with $T_e$ are somewhat balanced. Experimentally, saturation in energy relaxation was observed in Si-doped bulk GaN \cite{Liberis:2006} and a strained AlGaN/GaN heterostructure \cite{Matulionis:2005} . Experimental results \cite{Liberis:2009,Matulionis:2008} indicate that the high temperature side has phonon life-times $\tau_p$ one order of magnitude shorter than 1 ps. Using $\tau_p$=0.1 ps reduces the relaxation time $\tau_E$ to $\sim$0.12 ps for the static screening case (thick dashed line in Fig.~\ref{fig7}) and to $\sim$0.09 ps when dynamic screening is accounted for (thick dotted line in Fig.~\ref{fig7}), as the  hot phonon occupation numbers are reduced in the phonon re-absorption processes compared to the case of the longer lifetime of 1 ps. This rapid relaxation means no bottleneck for the power dissipation. Our calculated value $\sim$0.09 ps is nearly equal to the measured high-temperature relaxation time of 0.09 ps \cite{Matulionis:2008}. We note that in this case, despite it being weakened, the hot-phonon reabsorption should be included to obtain the relaxation time $\sim$0.09 ps, as we found that without HPE the relaxation would be faster with $\tau_E\approx$0.06 ps.

At high electron temperatures the measured relaxation time for the investigated lattice-matched heterostructure was found to decrease slowly with the electron temperature \cite{Matulionis:2008}. Our calculation shows that
the  one order of magnitude shorter phonon lifetime has reduced the relaxation time $\tau_E$ by only 0.3 ps  when static or dynamic screening is accounted for,
and therefore the high temperature relaxation time decreases slowly with $T_e$, in good agreement with experiment. These results also support the experimental finding \cite{Liberis:2009} that  the polar optical phonons have a shorter lifetime at a higher electron temperature, otherwise saturation in energy relaxation would occur, similar to that in bulk GaN \cite{Liberis:2006,Zhang:2014}.

We make a comparison of the energy relaxation in bulk GaN \cite{Liberis:2006,Zhang:2014} and the heterostructure. When hot phonons are ignored, the electron power loss is much greater in bulk GaN, which is  approximately three  times the power dissipation in the heterostructure when static screening is included. This is largely because the electron density of states is much higher in bulk than in the heterostructure.
However we found that the hot phonons play an important role in determining the high-temperature energy relaxation. For Si-doped bulk GaN with a volume electron density $10^{18}$ cm$^{-3}$, the high-$T_e$ relaxation time is around 0.2 ps [Fig.~7(a) of Ref.\cite{Zhang:2014}] with phonon lifetime 10 ps, which is longer than the relaxation time of $\sim$0.1 ps in the lattice-matched heterostructure.  With a higher electron density, $10^{19}$ cm$^{-3}$,  for instance, in bulk GaN the electron energy relaxation is found to be much slower due to the combined screening and hot-phonon effect \cite{Zhang:2014}. Therefore, the rapid energy relaxation with $\tau_E$ around 0.1 ps in the {\it heterostructure} means an efficient heat transfer from the hot electron gas to the lattice, which provides the heterostructure with an advantage to use in HFET devices.

\section{Conclusions}

In conclusion, we have studied energy relaxation for hot electrons in the quasi-2D channel of lattice-matched InAlN/AlN/GaN heterostructures using the DC and 3DP models.
The temperature of the quasi-2D electron gas in the narrow 44-$\AA$ channel can reach above 2500 K due to high electric power, much higher than the lattice temperature (room temperature).  In this study therefore non-equilibrium polar optical  phonons as well as electron degeneracy and screening from the mobile electrons
are taken into account.
Particular attention is paid to the effects of the two phonon models on the hot-electron
relaxation process in the GaN heterostructures. We calculated the electron temperature dependences of the electron power dissipation and energy relaxation time using a variety of phonon lifetime values and examined the 3DP model by comparing the results calculated with the two phonon models.
 We found that the 3DP model yields very close results to the DC model: with no hot phonons or screening the power loss calculated from the 3DP model is 5\% smaller than the DC power dissipation, whereas slightly larger 3DP power loss (by less than 4\% with a phonon lifetime from 0.1 to 1 ps) is obtained throughout the electron temperature range after including both the HPE and screening. Very close results are obtained also for the energy relaxation time with the two phonon models (with a percent deviation smaller than 5\%). As the investigated heterostructure has a channel narrower than the usual GaN-based heterostructures, therefore the 3DP phonon model is generally a good approximation to use for the study of the energy relaxation in GaN-based heterostructures.
We found that our results in the pure phonon emission case are consistent with the sum rules given by Mori and Ando \cite{Mori} and by Register \cite{Register}.  The discrepancy between the DC and 3DP results is caused by how much the high energy interface phonons contribute to the energy relaxation: their contribution is enhanced  in the pure emission process but is dramatically reduced after including the HPE. Debye screening overestimates the high-$T_e$ energy relaxation time
by $\sim$0.03 ps compared to the dynamic screening model whereas
with dynamic screening included anti-screening occurs at low electron temperatures (below $\sim$800 K) due to the dispersion of the quasi-2D electron plasma frequency.
Our calculation with both phonon models has obtained a great fall in  energy relaxation time $\tau_E$ at low electron temperatures ($T_e<$ 750 K) and slow decrease at the high temperatures with the use of decreasing phonon lifetime with $T_e$. The calculated temperature dependence of the relaxation time and the high-temperature relaxation time $\sim$0.09 ps are in good agreement with experimental results.
We also compared the quasi-2D hot-electron relaxation with the electron relaxation in bulk GaN and found that the hot phonons play a key role in slowing down the high-$T_e$ electron relaxation for bulk ($\tau_E$ $\sim$ 0.2 ps).  For the heterostructures, in contrast, the rapid energy relaxation ($\tau_E$ $\sim$ 0.09 ps) and sub-picosecond phonon decay provide an advantage which
benefits electron transport in the HFET devices by efficiently cooling down the extremely hot electrons to increase the electron mobility.

\begin{acknowledgments}
JZ acknowledges support from the Natural Science Research Funds of Jilin University, and AD and BKR would like to thank the Office of Naval Research, U.S. for funding under the Grant N00014-09-1-0777 sponsored by Dr. Paul Maki.
\end{acknowledgments}

\appendix

\section{Electron power dissipation based on the DC and 3DP models and the sum rules for electron-phonon scattering}

As can be seen from Sec. III, $1/\tau_n(q)$ and $1/\tau_{q_z}(q)$ are key quantities in the calculation of the power dissipation. First we look at how to treat $1/\tau_n(q)$ due to interface phonon scattering [Eq.~(\ref{nuif})]. Express the matrix element $M_{n,\mathbf{q}}(\mathbf{k};\mathbf{k}+\mathbf{q})$ in terms of the electron and phonon envelope functions, and substitute it into Eq.~(\ref{nuif}).  Then replace $\sum_{\mathbf{k}}$ by $\frac{A}{(2\pi)^2}\int kdkd\theta$
to convert the summation over electron wavevector to a double integral. The integration over angle $\theta$ can be  performed analytically, reducing the double integral to an integral over $k$ only. After some algebraic manipulation by changing the variable of integration, we find that the final $1/\tau_n(q)$ can be expressed in a simple form as
\begin{equation}
\frac{1}{\tau_n(q)}=\frac{m^*}{2\pi\hbar^3}\gamma_n^2\frac{1}{q}\lvert \Gamma_I(q)\rvert^2\sqrt{\pi}\left(\frac{k_BT_e}{\varepsilon_q}\right)^{1/2}\left[F_{-\frac{1}{2}}(\xi_{nq})-F_{-\frac{1}{2}}(\zeta_{nq})\right],
\label{nunqa1}
\end{equation}
where $\varepsilon_q=\hbar^2q^2/(2m^*)$,
$\Gamma_I(q)$ is the electron-interface-phonon overlap integral
\begin{equation}
\Gamma_I(q)=\int_{0}^{\infty}\phi^*(z)e^{-qz}\phi(z)dz,
\label{Iqif1}
\end{equation}
and $F_{-\frac{1}{2}}(y)$ is the complete Fermi-Dirac integral of order $-1/2$,
\begin{equation}
F_{-\frac{1}{2}}(y)=\frac{1}{\sqrt{\pi}}\int_{0}^{\infty}\frac{x^{-1/2}}{1+e^{x-y}}dx.
\label{fdm12}
\end{equation}
$\xi_{nq}$ and $\zeta_{nq}$ are two dimensionless quantities, given by
\begin{equation}
\xi_{nq}=(E_F-\Delta_{nq})/(k_BT_e), ~~~ \zeta_{nq}=\xi_{nq}-\hbar\omega_n/(k_BT_e),
\end{equation}
where $\Delta_{nq}$ has the dimension of energy,
\begin{equation}
\Delta_{nq}=(\varepsilon_q-\hbar\omega_n)^2/(4\varepsilon_q).
\end{equation}

Similarly,  one can obtain from Eq.~(\ref{nuhs}) the $1/\tau_{q_z}(q)$ expression for half-space phonon scattering,
\begin{equation}
\frac{1}{\tau_{q_z}(q)}=\frac{m^*}{2\pi\hbar^3}\gamma_{LO}^2\frac{1}{L_2}\frac{1}{q^2+q_z^2}\lvert \Gamma_H(q_z)\rvert^2\sqrt{\pi}\left(\frac{k_BT_e}{\varepsilon_q}\right)^{1/2}\left[F_{-\frac{1}{2}}(\xi_{q})-F_{-\frac{1}{2}}(\zeta_{q})\right],
\label{nuzqa1}
\end{equation}
where $\Gamma_H(q_z)$ is the electron-half-space-phonon overlap integral
\begin{equation}
\Gamma_H(q_z)=\int_{0}^{\infty}\phi^*(z)2\sin(q_zz)\phi(z)dz,
\label{Ihqz}
\end{equation}
and $\xi_{q}$ and $\zeta_{q}$ are given by
\begin{equation}
\xi_{q}=(E_F-\Delta_{q})/(k_BT_e),~~~ \zeta_{q}=\xi_{q}-\hbar\omega_{LO}/(k_BT_e),
\label{xizths}
\end{equation}
with the energy $\Delta_{q}$ being defined by
\begin{equation}
\Delta_{q}=(\varepsilon_q-\hbar\omega_{LO})^2/(4\varepsilon_q).
\label{epsqhs}
\end{equation}

For bulk LO phonon scattering, the generation rate $1/\tau(\mathbf{Q})$ is given by
\begin{equation}
\frac{1}{\tau(\mathbf{Q})}=\frac{m^*}{\pi\hbar^3}\gamma_{LO}^2\frac{1}{L_1+L_2}\frac{1}{q^2+q_z^2}\lvert \Gamma_B(q_z)\rvert^2\sqrt{\pi}\left(\frac{k_BT_e}{\varepsilon_q}\right)^{1/2}\left[F_{-\frac{1}{2}}(\xi_{q})-F_{-\frac{1}{2}}(\zeta_{q})\right],
\label{nuQa1}
\end{equation}
where $\Gamma_B(q_z)$ is the electron-LO-phonon overlap integral
\begin{equation}
\Gamma_B(q_z)=\int_{0}^{\infty}\phi^*(z)e^{-iq_zz}\phi(z)dz.
\label{IBu}
\end{equation}

In what follows we confine ourselves to a discusion of the sum rules so we ignore screening and hot phonons. We recall that the calculation of the electron {\it power dissipation} involves summation over $q_z$ for both half-space and bulk phonon modes [Eq.~(\ref{pdDC}) and (\ref{pd3DP})]. Inserting the $\Gamma_H(q_z)$ expression Eq.~(\ref{Ihqz}) in Eq.~(\ref{nuzqa1}) and then performing summation over $q_z$ for the $q_z$-dependent factor, we obtain
\begin{equation}
\sum_{q_z>0}\frac{1}{L_2}\frac{1}{q^2+q_z^2}\lvert \Gamma_H(q_z)\rvert^2=\frac{1}{q}F_H(q),
\label{sumqzhs}
\end{equation}
where $F_H(q)$ is the form factor for half-space phonons as defined by Mori and Ando \cite{Mori},
\begin{equation}
F_H(q)=\int_0^{\infty} dz\int_0^{\infty} dz'\phi^*(z)\phi(z)\left(e^{-q\lvert z-z'\rvert}-e^{-q\lvert z+z'\rvert}\right)\phi^*(z')\phi(z').
\label{formFH}
\end{equation}
Note that in obtaining Eq.~(\ref{sumqzhs}) we have used the integration formula
\begin{equation}
\int_0^{\infty}\frac{\cos ax}{\beta^2+x^2}dx=\frac{\pi}{2\beta}e^{-\beta\lvert
a\rvert}~~~~~(\beta>0)~.
\end{equation}

For the 3D bulk modes [Eqs.~(\ref{nuQa1}) and (\ref{IBu})], similarly one finds
\begin{equation}
\sum_{q_z}\frac{1}{L_1+L_2}\frac{1}{q^2+q_z^2}\lvert \Gamma_B(q_z)\rvert^2=\frac{1}{2q}F_B(q),
\label{sumqz3D}
\end{equation}
with the form factor given by
\begin{equation}
F_B(q)=\int_0^{\infty} dz\int_0^{\infty} dz'\phi^*(z)\phi(z)e^{-q\lvert z-z'\rvert}\phi^*(z')\phi(z').
\label{formFB}
\end{equation}

For the interface modes [Eqs.~(\ref{nunqa1}) and (\ref{Iqif1})], it is readily found that the form factor $F_I(q)$ \cite{Mori} is simply the square of the overlap integral $\Gamma_I(q)$
\begin{equation}
F_I(q)=\left[
\int_{0}^{\infty}dz \phi^*(z)e^{-qz}\phi(z)\right]^2.
\label{formFI}
\end{equation}

Using Eqs.~(\ref{formFH}), (\ref{formFB}) and (\ref{formFI}) then we restore the form factor sum rule given by Mori and Ando \cite{Mori}
\begin{equation}
F_B(q)=F_H(q)+F_I(q).
\label{ruleF}
\end{equation}

We now proceed to applying this sum rule to the power dissipation as obtained from the DC and 3DP models. We first consider a special case, that is, when the quasi-2D channel in GaN is sufficiently wide that the form factor for the interface phonons $F_I(q)$
can be neglected. This leads to the removal of the contribution to the power dissipation from all the interface modes. Inserting the $1/\tau_{q_z}(q)$ and
$1/\tau(\mathbf{Q})$ expressions (\ref{nuzqa1}) and (\ref{nuQa1})   into the power dissipation equations (\ref{pdDC}) and (\ref{pd3DP}) respectively, and then using the obtained identities Eqs.~(\ref{sumqzhs}) and (\ref{sumqz3D}), we find that the two resulting power dissipation expressions given by the DC and 3DP models are {\it identical}. In the general case, of course one has $F_I(q)\ne 0$, the two phonon models do not yield the same power dissipation, and the difference is caused entirely by how much the true interface modes (eigenmodes) contribute compared to that when they are treated simply as bulk LO modes in terms of the phonon frequency and electron-phonon coupling strength. This can be made more clear by the following proof.
If we ignore the difference in the material parameters, such as the phonon frequencies and dielectric constants, of the two constituents of the heterostructure, and use only the GaN parameters, the interface mode frequency that is given by the solution of $\epsilon(\omega)+\bar{\epsilon}(\omega)=0$ simply reduces to the LO phonon frequency $\omega_{LO}$ of GaN, and $\frac{1}{\epsilon_n}$ for the interface modes reduces to $\frac{1}{\epsilon_{LO}}$ for the bulk LO modes, thus making the coupling constant $\alpha_n$ for the interface modes, $\alpha_n=\frac{e^2}{\hbar}(\frac{m^*}{2\hbar\omega_n})^{1/2}/\epsilon_n$ \cite{Mori}, reduce to the Fr\"ohlich coupling constant $\alpha_{LO}$ for the bulk LO modes,   $\alpha_{LO}=\frac{e^2}{\hbar}(\frac{m^*}{2\hbar\omega_{LO}})^{1/2}/\epsilon_{LO}$. This is exactly the scenario that results from the degenerate modes approximation \cite{Register,Mori}.
It then follows that the use of the Mori-Ando sum rule [Eq.~(\ref{ruleF})]
results in the same DC and 3DP power dissipation.
Furthermore, the same electron-phonon scattering rate is also obtained
 with the two phonon models, which is consistent with the sum rule for the electron-phonon interaction given by Register \cite{Register}.
Of course the two constituent materials are different, and the true interface modes are not bulk GaN LO modes, and therefore in general the DC power dissipation is different from that evaluated with the 3DP model, even in the non-HPE case, as our numerical results in Sec. IV have shown.

\section{Phonon wavevector dependences of energy $\Delta_q$ and the Fermi-Dirac integral  $F_{-\frac{1}{2}}(\xi_q)$}

The following illustration is made for the half-space modes but it can be equally applied to the interface and bulk LO modes as well. The energy expression appearing in the delta function of Eq.~(\ref{nuhs}) indicates energy conservation when an electron at state
$\mathbf{k}$ makes a transition up to state $\mathbf{k}+\mathbf{q}$ after a half-space phonon with wavevector $\mathbf{q}$ is absorbed. If $\theta$ is the angle between $\mathbf{k}$ and $\mathbf{q}$, one finds $\cos\theta=\frac{\hbar\omega_{LO}-\varepsilon_q}{\hbar^2kq/m^*}$. As $\lvert\cos\theta\rvert\leq$ 1, one obtains $E_k\geq\Delta_q$ [$\Delta_q$ is given by Eq.~(\ref{epsqhs})].  Therefore, the energy $\Delta_q$, which arises from energy conservation plus momentum  conservation, has a clear physical meaning; that is,  given a phonon wavevector, $\Delta_q$ is the {\it minimum} kinetic energy of the electron for the phonon absorption to occur. Fig.~\ref{fig9}(a) shows the energy $\Delta_q$ as a function of the wavevector $q$, where for simplicity $\Delta_q$ and $q$ are made dimensionless with respect to the LO phonon energy $\hbar\omega_{LO}$ and electron wavevector for threshold LO phonon emission $k_0$, respectively. We see that the energy $\Delta_q$
decreases steeply to the minimum value of 0 at $q=k_0$, and then slowly increases as wavevector $q$ becomes larger.
Eq.~(\ref{nuzqa1}) shows that the generation rate $1/\tau_{q_z}(q)$  is proportional to the difference of the Fermi-Dirac integrals at the two arguments differing by only $\hbar\omega_{LO}/(k_BT_e)$ [also refer to Eq.~(\ref{xizths})], owing to scattering by half-space phonons of energy $\hbar\omega_{LO}$. Further, the Fermi-Dirac integral $F_{-\frac{1}{2}}(y)$ [Eq.~(\ref{fdm12})] is a monotonically increasing function, and this ensures that $F_{-\frac{1}{2}}(\xi_{q})-F_{-\frac{1}{2}}(\zeta_{q})$ and consequently the rate $1/\tau_{q_z}(q)$  is always positive as required physically. Therefore,
knowing the dependence of $F_{-\frac{1}{2}}(\xi_q)$ (or $F_{-\frac{1}{2}}(\zeta_{q})$ equally) on the wavevector $q$ is fundamental to understanding the $q$-dependence of the non-equilibrium phonon occupation number in Sec. IV.  As shown in Fig.~\ref{fig9}(b) (solid line), $F_{-\frac{1}{2}}(\xi_q)$ increases rapidly from 0 to the maximum value at $q=k_0$, and then decreases with further increasing $q$. In both limits of $q\rightarrow 0$ and $q\rightarrow\infty$ the Fermi-Dirac integral is zero leading to  null phonon generation.
In the non-degenerate Boltzmann approximation, one has $F_{-\frac{1}{2}}(y)=e^y$,  making $F_{-\frac{1}{2}}(\xi_q)=e^{\xi_q}$ simply proportional to $e^{-\Delta_q/k_BT_e}$. Then one finds that
$F_{-\frac{1}{2}}(\xi_q)$ depends on $q$ according to $e^{-\alpha(\frac{k_0}{q})^2/4}$ for $q\rightarrow 0$, and according to $e^{-\alpha(\frac{q}{k_0})^2/4}$ for $q\rightarrow \infty$ where $\alpha=\hbar\omega_{LO}/(k_BT_e)$.
For an electron density of $7\times 10^{11}$/cm$^2$
at the electron temperature $T_e=2\hbar\omega_{LO}/k_B$=2110 K, for instance,
the non-degeneracy approximation $e^{\xi_q}$ and the Fermi-Dirac integral $F_{-\frac{1}{2}}(\xi_q)$ are very close, making the lower two curves in Fig.~\ref{fig9}(b) coincide. In typical GaN based heterostructures the electron density is quite high. For electron density $1.2\times 10^{13}$/cm$^2$, the non-degeneracy approximation (dot-dashed line)
is quite large with a peak value of 60\% larger than the maximum value of  $F_{-\frac{1}{2}}(\xi_q)$ (solid line).
 Nevertheless the dependence of $F_{-\frac{1}{2}}(\xi_q)$ on the phonon wavevector $q$ governs the $q$-dependence of the non-equilibrium phonon occupation number, as shown in Sec. IV.


\newpage

\begin{figure}
\caption
{(Color online) (a) Non-equilibrium phonon occupation numbers vs in-plane phonon wavevector of the half-space phonons for a number of $q_z$ indices as labeled ($q_z=l\pi/L_2$, $l\geq 1$, $L_2$ being the dimension of the GaN half-space), generated from energy relaxation of the hot electrons at temperature 1000 K in a typical lattice-matched heterostructure In$_{0.18}$Al$_{0.82}$N/AlN/GaN with an areal electron density of  $1.2\times 10^{13}$/cm$^2$ (corresponding to an effective channel width of 44 \AA), calculated with a polar optical phonon life-time $\tau_p$=0.1 ps. The horizontal dotted line shows the thermal equilibrium occupation number at room temperature, $N(\omega_{LO},T_0)=0.03$.  
 (b) Phonon-wavevector-dependent factor, $\Gamma_H^2(q_z)k_0^2/(q^2+q_z^2)$, of the squared electron-phonon interaction matrix elements of half-space modes, as functions of $q_z$ for three in-plane wavevectors $q$=0.5, 1, 2$k_0$, where $k_0$ is the characteristic electron wave-vector for threshold LO phonon emission, $k_0=\sqrt{2m^*\omega_{LO}/\hbar}$.
}
\label{fig1}
\vspace*{5mm}

\caption
{(Color online) Non-equilibrium phonon occupation numbers of the lower-energy ($IF_1$) and higher-energy ($IF_2$) interface phonons from energy relaxation of the hot electrons at temperature 1000 K in the lattice-matched In$_{0.18}$Al$_{0.82}$N/AlN/GaN  heterostructure,  calculated with a phonon life-time $\tau_p$=0.1 ps for (a) excluding and (b) including screening with the static (Debye) and dynamic screening models.
}
\label{fig2}
\vspace*{5mm}

\caption
{
(Color online) Average power dissipated per electron vs electron temperature for (a) a strained Al$_{0.05}$Ga$_{0.95}$N/GaN heterostructure with an effective GaN-channel width of
110 $\AA$ (corresponding to an areal electron density of $7\times 10^{11}$/cm$^2$) and (b) the lattice-matched In$_{0.18}$Al$_{0.82}$N/AlN/GaN heterostructure with an effective GaN-channel width of 44 $\AA$  (corresponding to the areal electron density of  $1.2\times 10^{13}$/cm$^2$), calculated with the dielectric continuum (DC) and three-dimensional phonon (3DP) models for excluding or including the hot-phonon effect (HPE) with two phonon life-time values $\tau_p$=0.5 and 2 ps as labeled.
}
\label{fig3}
\vspace*{5mm}
\end{figure}

\newpage

\begin{figure}
\caption
{
(Color online) (a) Average power dissipated per electron vs electron temperature in the lattice-matched In$_{0.18}$Al$_{0.82}$N/AlN/GaN heterostructure, due to scattering with the half-space (HS), lower-energy ($IF_1$) and higher-energy ($IF_2$) interface phonons, respectively,
and (b) higher-energy interface phonon ($IF_2$) generation number per unit time $W_2(q)$ [Eq.~(\ref{genif})]  vs in-plane phonon wave-vector at the electron temperature 2500 K,
when the hot-phonon effect (HPE) is excluded or included with a phonon life-time of $\tau_p$=0.5 ps.
In (b) the phonon generation $W_2(q)$ values of including the hot phonons have been enlarged by ten times.
}
\label{fig4}
\vspace*{5mm}

\caption
{(Color online) Average power dissipated per electron vs electron temperature in the lattice-matched In$_{0.18}$Al$_{0.82}$N/AlN/GaN heterostructure, calculated with the dielectric continuum (DC) and three-dimensional phonon (3DP) models for the six cases as labeled, namely, (i) without hot-phonon effect (HPE) or  screening (scr), (ii) including only the HPE, (iii) including only static (Debye) screening, (iv) including only dynamic screening, (v) including both the HPE and static screening,
 and (vi) including both the HPE and dynamic screening. The polar optical phonon life-time of 1 ps is used when the hot phonons are included for the three cases (ii), (v), (vi).
}
\label{fig5}
\vspace*{5mm}

\caption
{(Color online) Reduction factors versus electron temperature associated with only the hot-phonon effect (HPE), only screening (Debye screening or dynamic screening), and both HPE and screening calculated with the dielectric continuum (DC) and three-dimensional phonon (3DP) models as labeled, where a polar optical phonon lifetime of 1 ps is used for hot phonons in the lattice-matched In$_{0.18}$Al$_{0.82}$N/AlN/GaN heterostructure. Also drawn is a dotted horizontal line $\beta$=1, the part of curves below which indicates antiscreening.  
}
\label{fig6}
\vspace*{5mm}

\end{figure}

\begin{figure}

\caption
{(Color online) Energy relaxation times vs electron temperature of the hot electrons in the lattice-matched In$_{0.18}$Al$_{0.82}$N/AlN/GaN heterostructure, calculated with the dielectric continuum (DC) and three-dimensional phonon (3DP)  models for several cases as labeled: with no hot-phonon effect (HPE) or screening (scr), including both  HPE and static (Debye) screening,
 and including both the HPE and dynamic screening with a phonon lifetime of $\tau_p$=1 ps. The DC results at high electron temperatures using both screening models calculated with the phonon lifetime 0.1 ps are also shown.
}
\label{fig7}
\vspace*{5mm}

\caption
{(Color online) Percent deviations of the energy relaxation times evaluated with the three-dimensional phonon (3DP) approximation with respect to the dielectric continuum (DC) calculation as functions of the electron temperature in the In$_{0.18}$Al$_{0.82}$N/AlN/GaN heterostructure for several cases as indicated: with no hot-phonon effect (HPE) or screening (scr), including both the HPE and static  (Debye) screening,
 and including both the HPE and dynamic screening with a phonon lifetime of $\tau_p$=1 ps. The relative deviations at high electron temperatures  using both screening models calculated with the phonon lifetime 0.1 ps are also shown.
}
\label{fig8}
\vspace*{5mm}

\caption
{(Color online) (a) The energy $\Delta_q$, namely, the {\it minimum} electron kinetic energy for phonon absorption, given by $\frac{\Delta_q}{\hbar\omega_{LO}}=\frac{[(\frac{q}{k_0})^2-1]^2}{4(\frac{q}{k_0})^2}$, as a function of the phonon wavevector $q$; (b) the Fermi-Dirac integral $F_{-\frac{1}{2}}(\xi_q)$, where $\xi_{q}=(E_F-\Delta_{q})/(k_BT_e)$, as functions of the phonon wavevector $q$ for the two electron densities $1.2\times 10^{13}$ (solid) and $7\times 10^{11}$ cm$^{-2}$ (dotted) at the same electron temperature $T_e=2\hbar\omega_{LO}/k_B$=2110 K. Also shown is the non-degenerate Boltzmann approximation,
$e^{\xi_q}$, for the Fermi-Dirac integral $F_{-\frac{1}{2}}(\xi_q)$ for the two electron densities respectively (dot-dashed and dashed). Note that $k_0$ is the characteristic electron wave-vector for threshold LO phonon emission, $k_0=\sqrt{2m^*\omega_{LO}/\hbar}$.
}
\label{fig9}
\vspace*{5mm}

\end{figure}

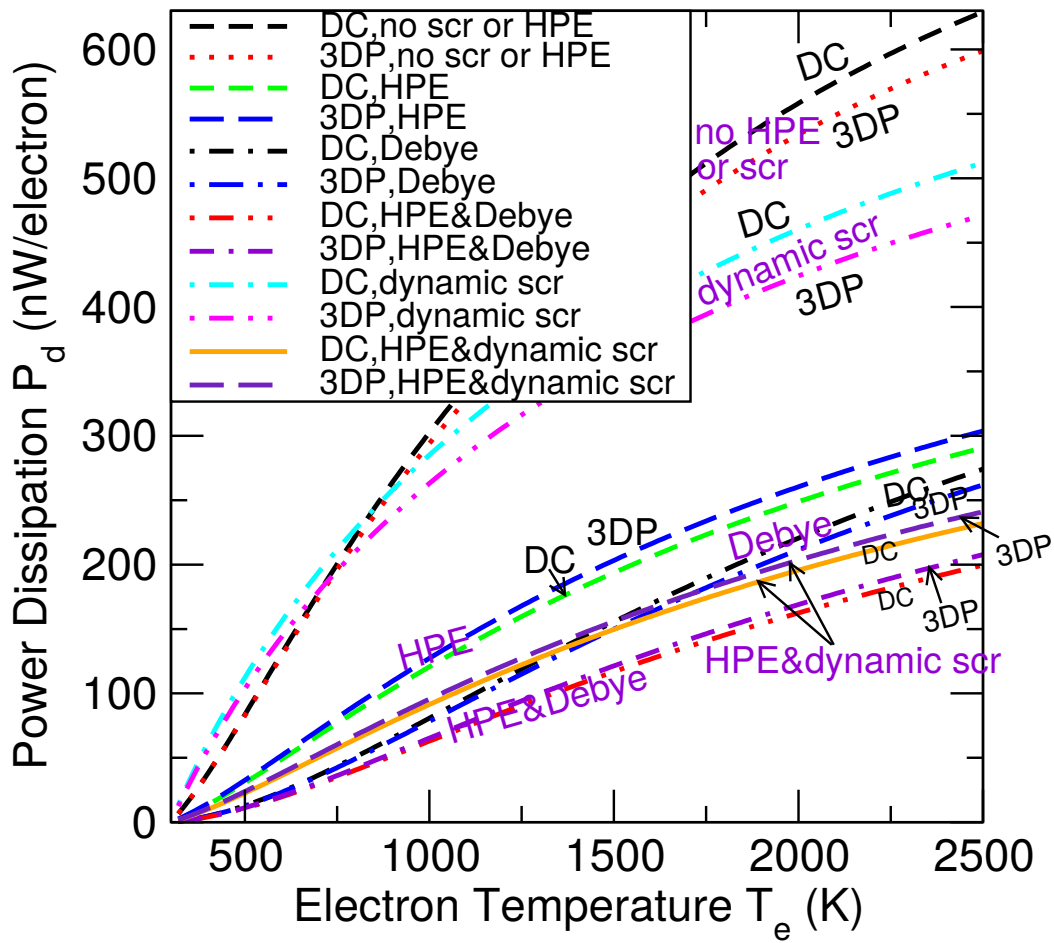
\end{document}